\documentclass{SciPost}

\usepackage[utf8]{inputenc} 
\usepackage[T1]{fontenc} 	
\usepackage[english]{babel} 

\usepackage[bitstream-charter]{mathdesign}
\usepackage{geometry} 		
\usepackage{amsmath} 		
\usepackage{mathtools} 		
\usepackage{float} 			
\usepackage{graphicx} 		
\usepackage{tabularx} 		
\usepackage{booktabs} 		
\usepackage{color, xcolor} 	
\usepackage{pdfpages} 		
\usepackage{extarrows} 		
\usepackage{multirow} 		
\usepackage{multicol} 		
\usepackage{enumitem} 		
\usepackage{xspace} 		
\usepackage{stackrel} 		
\usepackage{tikz} 			
\usepackage{braket} 		
\usepackage{bm} 			
\usepackage{tensor} 		
\usepackage{slashed} 		
\usepackage{siunitx} 		
\usepackage{lastpage} 		
\usepackage{cite} 			
\usepackage[normalem]{ulem} 
\usepackage{fontawesome} 	
\usepackage{tocloft} 		
\usepackage{titlesec} 		
\usepackage{doi} 			
\usepackage{hyperref} 		
\usepackage[most]{tcolorbox} 					
\usepackage[nameinlink, capitalize]{cleveref} 	
\usepackage[nottoc, notlot, notlof]{tocbibind} 	
\usepackage[ruled, vlined]{algorithm2e} 		
\usepackage{makecell}
\usepackage[makeroom]{cancel}
\usepackage{feynmf}

\makeatletter
\def\BState{\State\hskip-\ALG@thistlm}
\makeatother

\makeatletter
\@ifundefined{pdfoutput}{}{\DeclareGraphicsRule{*}{mps}{*}{}}
\makeatother

\makeatletter
\DeclareRobustCommand*{\bfseries}{%
   \not@math@alphabet\bfseries\mathbf
   \fontseries\bfdefault\selectfont
   \boldmath
}
\makeatother

\hypersetup{
	colorlinks=true, 			
	linkcolor={red!50!black}, 	
	citecolor={blue!50!black}, 	
	urlcolor={blue!80!black} 	
} 

\DeclareSymbolFont{usualmathcal}{OMS}{cmsy}{m}{n}
\DeclareSymbolFontAlphabet{\mathcal}{usualmathcal}

\SetArgSty{textnormal}
\SetKwComment{Comment}{{\small\#}~}{}
\SetCommentSty{mycommfont}

\setitemize{itemsep=0pt, parsep=0pt} 				
\setenumerate{itemsep=0pt, parsep=0pt} 				
\setitemize{itemsep=2pt,topsep=2pt,parsep=0pt,partopsep=0pt,leftmargin=*}
\setenumerate{itemsep=0pt,topsep=2pt,parsep=0pt,partopsep=0pt,labelindent=3pt,leftmargin=*}
\usepackage{amsmath}
 
\usepackage{amsthm} 		
\theoremstyle{definition}

\fancypagestyle{SPstyle}{
\fancyhf{}
\lhead{\colorbox{scipostblue}{\bf \color{white} ~SciPost Physics Community Reports }}
\rhead{{\bf \color{scipostdeepblue} ~Submission }}

\fancyfoot[C]{\textbf{\thepage}}
}

\newcommand{\gen}{\text{gen}}
\newcommand{\data}{\text{data}}

\newcommand{\pgen}{p_\gen}
\newcommand{\Ngen}{n_\gen}
\newcommand{\Dgen}{D_\gen}
\newcommand{\pdata}{p_\data}

\newcommand{\Ddata}{D_\data}

\newcommand{\Ntrain}{n_\text{train}}

\newcommand{\Nequiv}{n_\text{equiv}}

\begin{document}

\pagestyle{SPstyle}


\begin{center}{\Large \textbf{Generative Models and Statistical Validation 
 }}\end{center}
\begin{center}
    {Part of the VERaiPHY Initiative}
\end{center}
\begin{center}
Sascha Diefenbacher\textsuperscript{*,1,2},
Sofia Palacios Schweitzer\textsuperscript{*,1,3} and
Gregor Kasieczka\textsuperscript{$\circ$,4}

\end{center}

\begin{center}
{\bf 1} Institut für Theoretische Physik, Universität Heidelberg, Germany \\
{\bf 2} Physics Division, Lawrence Berkeley National Laboratory, Berkeley, USA \\
{\bf 3} NHETC, Department of Physics \& Astronomy, Rutgers University, Piscataway, NJ, USA\\
{\bf 4} Institut für Experimentalphysik, Universität Hamburg, Germany \\

\vspace{1.5mm}

{\bf*} Leading authors \qquad {\bf$\circ$} Advisor\\

\end{center}

\begin{center}
\today
\end{center}


\section*{Abstract}
{\bf Generative machine learning has become an essential tool in theoretical and experimental physics, especially in the context of fast surrogates and density estimators. In this work, we first introduce the underlying framework of modern generative networks and then discuss challenges in quantifying their accuracy, precision, and statistical power.}

\vspace{10pt}
\noindent\rule{\textwidth}{1pt}
\tableofcontents\thispagestyle{fancy}
\noindent\rule{\textwidth}{1pt}
\vspace{10pt}
\clearpage

\section{Introduction}
\label{sec:intro}
Generative machine learning is the key engine behind revolutionary techniques such as large language models and image and video generators. In the last five years, state-of-the-art network architectures designed to model language or images have reached a level of precision that has promoted their use in simulation and inference tasks across a wide range of scientific subfields. Fields in fundamental physics that require large amounts of precise simulations and robust analysis tools for high-dimensional data analysis, are natural candidates for the use of generative machine learning. 
However, the adoption of generative machine learning in physics confronts a fundamental tension. Most of the time, classical simulations of a physical system are based on first-principles calculations. We write down a Lagrangian, derive the dynamics, and propagate these dynamics numerically. The related uncertainties are controllable. We can either increase statistics, or improve the underlying theory. Generative networks operate differently. They learn to approximate a target distribution from a finite training sample, without explicit access to the underlying physical laws. This empirical foundation raises questions that, while present in any simulation, become considerably harder to answer:
\begin{enumerate}
    \item Does the learned distribution faithfully represent the underlying true distribution, or does the network introduce systematic distortions? In classical simulations, shortcomings can often be anticipated and related to specific assumptions or approximations in the theoretical formulation. For generative networks this is no longer possible as there is no built-in diagnostic that tells us what the model is missing. How do we detect and quantify such mismodeling?
    \item A model may be accurate on average yet overconfident or underconfident in its predictions. How do we quantify uncertainties arising from finite training data and residual mismodeling, ensure that they are well-calibrated, and propagate them to downstream analyses?
    \item Many applications aim to augment limited simulation samples by generating additional synthetic events. Under what conditions can we reliably generate statistics beyond our training sample, and when does such amplification become self-deception?
\end{enumerate}
These questions are not unique to physics, but fundamental physics is one of the few domains where we have access to samples of a meaningful ground truth distribution and require rigorous statistical standards. For example, in language modeling it is difficult to even define accuracy or uncertainty as there is no first-principles theory of text.
In physics, the situation is different. Simulations enter directly into measurements, they define analysis selections, model backgrounds, and propagate into systematic uncertainties. A generative model that replaces or augments such simulations inherits these responsibilities. Beyond surrogate modeling, generative networks also enable new high-dimensional analysis techniques that require the same rigor. 
This has motivated substantial methodological development within the physics community, and in this work, we give an overview of current approaches.
We start by introducing a general mathematical formalism of generative machine learning networks in Sec.~\ref{sec:general}, followed by exemplary use-cases of generative machine learning in fundamental physics in Sec.~\ref{sec:use_cases}. In Sec.~\ref{sec:validation}, we describe ways to quantify the accuracy of generative networks and list sources of why a network might mismodel certain regions of the data space. 
To be a reliable tool, generative networks must be not only accurate but also precise, in the sense of producing well-calibrated uncertainties. How such uncertainties can be implemented into different algorithms is presented in Sec.~\ref{sec:Uncertainties}.
We discuss how ML-based data augmentation can be statistically interpreted in Sec.~\ref{sec:amplification}.
Finally, many of the methods discussed here remain under active development, with open questions and unresolved limitations. We try to contextualize these in the conclusion and outlook in Sec.~\ref{sec:conclusion}.

This article contributes to VERaiPHY (Validation \& Evaluation for Robust AI in PHYsics), a PHYSTAT review series establishing verification and validation standards for machine learning across particle physics, astrophysics, and cosmology.

\section{What are Generative Networks?}
\label{sec:general}
\subsection{Learning a Transformation}
Generative networks aim to sample from a simple latent distribution, typically Gaussian noise, and transform these samples to represent data drawn from an underlying physical (and sometimes unknown) data distribution. This process is usually modeled as a learnable transformation. If we can construct a function that maps Gaussian noise to physically meaningful samples, we have effectively captured the target distribution. Such a transformation needs to respect the following criteria:
\begin{enumerate}
    \item if a point of the latent space is fed into the transformation, the output should represent a data point drawn from the target data distribution
    \item if we repeatedly sample from the latent space and pass it through the learnable transformation, the resulting distribution should approximate the target data distribution
    \item training and inference time should be feasible. 
\end{enumerate}
The main challenge lies in defining a suitable loss function that, when optimized, ensures complete and accurate coverage of the target distribution. One of the earliest types of generative networks,
Generative Adversarial Networks (GANs)~\cite{gans}, addresses this by learning such a transformation using neural networks. A so-called generator $G_\theta(x)$  learns the mapping between Gaussian noise and physical data 
\begin{align}
    z \sim p(z)=\mathcal{N} (0,1) \xrightarrow[]{\quad G_\theta(z) \quad} x \sim p_\theta(x) 
\end{align}
by utilizing an additional discriminator (classifier). This discriminator is trained to distinguish between newly generated samples and real samples from the training data. The mapping from latent space to physical space is encoded directly in a neural network that can, in principle, map between spaces of different dimensionalities. The network learns to produce samples that are increasingly indistinguishable from real data over the course of training. Under the assumption of a perfect discriminator and infinite training data, a GAN should map out the entire target distribution. In practice, it is considered a common shortcoming of GANs that they suffer from mode collapse, where the generator learns to produce samples from only a subset of the target distribution while failing to cover other regions.\\
The variational autoencoder (VAE) \cite{vae} underlies a different approach, although again two functions are learned: an encoder $E_\theta(x)$, which maps samples from the physical space $x$ to samples of the Gaussian latent space $z$, and a decoder $D_\theta(z)$, which maps back from the latent space to the physical space. As for GANs, the dimensionality of $x$ and of $z$ does not have to match, instead VAEs often encode a dimensional compression or expansion. During training, the model enforces the latent space to be Gaussian distributed while minimizing the difference between the original sample and its reconstruction, encouraging the decoder to learn a transformation that produces physically plausible samples. Once training converges, the decoder alone can be used to generate new data. By sampling from the Gaussian latent space and passing those samples through the decoder network, we obtain synthetic data that approximates the true physical distribution
\begin{align}
    x \sim p(x) \xrightarrow[]{\quad E_\theta(x)\quad } z \sim p(z) = \mathcal{N}(0,1) \xrightarrow[]{\quad D_\theta(z)\quad} x \sim p_\theta(x) \; .
\end{align}
Invertible Neural Networks (INNs), also known as normalizing flows~\cite{INNs}, build directly on the VAE concept but with a crucial constraint. Rather than learning separate encoder and decoder networks, an INN constructs a single transformation $f_\theta(x)$ that is bijective by design. The decoder is then simply the mathematical inverse $f^{-1}_\theta(z)$, not a separate numerical approximation. 
\begin{align}
   x \sim p(x) \xrightarrow[]{\quad f_\theta(x) \quad } z \sim p(z) = \mathcal{N}(0,1) \xrightarrow[]{\quad f_\theta^{-1}(z) \quad }x\sim p_\theta(x)
\end{align}
Due to its invertibility, $f_\theta(x)$ provides a closed form expression to estimate the underlying density $p(x)$ via the change of variable formula
\begin{align}
    \label{eq:change_of_variable}
    p_\theta(x) dx = p(z) dz \qquad \longrightarrow \qquad
    p_\theta(x) = p(f_\theta(x)) \det \left|\frac{\partial f_\theta(x)}{dx} \right| \; ,
\end{align}
where $p(z)$ has to be a tractable latent space distribution like $\mathcal{N}(0,1)$. This property makes INNs particularly attractive for inference tasks in physics as neither VAEs nor GANs are suitable for an exact density estimation. At the same time, Eq.~\eqref{eq:change_of_variable} provides an ideal training objective to generate new samples 
\begin{align}
    x = f_\theta^{-1}(z) \sim p_\theta(x) \;,
\end{align}
that maximize the likelihood $p_\theta(x)$. 
The architectural constraint of having an invertible mapping $f_\theta(x)$ means that we cannot use arbitrary neural networks for it.
Instead, INNs are constructed by composing simple bijective building blocks whose Jacobian determinants are tractable to compute. Neural networks will then parameterize components within these bijectors rather than the full transformation. Expressiveness can be achieved through depth. Stacking many simple bijective layers yields a flexible overall mapping. Nevertheless, the bijectivity constraint imposes fundamental limits on the class of distributions that can be efficiently represented. In addition, requiring bijectivity also means that the dimensionality of the latent space must match the dimensionality of the data space of interest, in contrast to a VAE, where a dimensional compression is often desired. \\\\
Diffusion models~\cite{Diffusion, DDPMs} pursue the same goal of inverting the mapping between noise and data, but take a fundamentally different route. Rather than constructing a deterministic bijection, they describe the mapping between physical data and noise as a continuous stochastic process. In the forward process, a data point $x_0 \equiv x \sim p(x, t=0)$ follows a stochastic trajectory from $t \in [0,1]$ until mapped to a latent space point $x_1 \equiv x \sim p(x, t=1) = \mathcal{N}(0,1)$. This forward mapping can be described by a stochastic differential equation (SDE)
\begin{align}
    \label{eq:forward_SDE}
    dx = f(x,t) dt + g(t) \; dW_t \;,
\end{align}
where $f(x,t)$ is referred to as the drift term, $g(t)$ the global time-dependent diffusion term and $dW_t$ a noise infinitesimal. As this forward process is fixed and requires no learning, $f(x,t)$ and $g(t)$ are chosen, such that the forward pass becomes tractable, by gradually adding Gaussian noise. The reverse process (noise to data) is related to a second SDE
\begin{align}
    \label{eq:backwards_SDE}
    dx = \left[f(x,t) - \frac{1}{2}g(t)^2 \nabla_x \log p(x,t)\right]dt + g(t) d\overline{W}_t 
\end{align}
going backwards in time, where $f(x,t)$ and $g(t)$ are determined by the forward process, $\overline{W}_t$ the noise infinitesimal of the backward process and the score $s(x,t) \equiv \nabla_x \log p(x,t)$ of the underlying density $p(x,t)$ must be learned by a neural network. In contrast to INNs, the network estimating $s_\theta(x,t) \approx s(x,t) $ can have an arbitrary architecture. Once converged, a sample $x_1 \sim p(x_1)$ can be mapped to $x_0 \sim p_\theta(x_0)$, by using Eq.~\eqref{eq:backwards_SDE} in combination with a SDE solver. Unlike INNs, where a single network evaluation produces a sample, diffusion models require integrating through many time steps, with a network evaluation at each step. The flow can be described by
\begin{align}
    x_0 \sim p(x_0) \xrightarrow[]{\quad p(x,t| x_0) \quad} x_1 \sim p(x_1) \xrightarrow{\quad p_\theta(x,t | x_1)\quad} x_0 \sim p_\theta(x_0) \;,
\end{align}
where $p(x,t | x_0)$ describes the known forward transition density for a given data point $x_0 \sim p(x_0)$ and $p_\theta(x,t|x_1)$ the learned reverse probability for a given latent space point $x_1 \sim p(x_1)$. Although we present the formalism as continuous in time, it is often discretized into a fixed number of time-steps~\cite{DDPMs}.\\
Similar to INNs, diffusion models provide exact likelihood expressions as the evolution of the probability density is governed by the Fokker-Planck equation
\begin{align}
    \label{eq:FPE}
    \frac{\partial p(x,t)}{\partial t} &= -\nabla_x \left[ f(x,t)\, p(x,t) \right] + \frac{1}{2} g(t)^2\, \nabla_x^2 p(x,t) \notag \\
    &= - \nabla_x \left[\left(f(x,t) - \frac{1}{2} g(t)^2 \nabla_x \log p(x,t) \right) p(x,t)\right]\;,
\end{align}
which relates the microscopic dynamics of the SDE $dx(t)$ to the macroscopic probability flow $p(x,t)$. However, this requires integrating along the trajectory, making an exact evaluation computationally intractable in practice.
The second expression of Eq.~\eqref{eq:FPE} allows us to relate the probability flow of a SDE to an ordinary differential equation (ODE) which shares the same marginals by constructing a velocity field
\begin{align}
    v(x,t) \equiv f(x,t) - \frac{1}{2} g(t)^2 \nabla_x p(x,t)\notag 
\end{align}
such that the trajectories take the deterministic form of
\begin{align}
    \label{eq:ODE}
    dx(t) = v(x,t) dt \;.
\end{align}
This methodology is often referred to as (conditional) flow matching~\cite{CFM}.
Instead of learning a score function, the velocity field can be parametrized and learned directly by a neural network $v_\theta(x,t)$. Once converged, a new sample $x_0 \sim p_\theta(x_0)$ can be generated by solving the ODE 
\begin{align}
    x_0 = x_1 - \int_0^1 dt \; v_\theta(x,t) \quad \text{with} \quad x_1 \sim p(x_1) \;.
\end{align}
Similarly, the exact likelihood can be computed by integrating the Fokker-Planck equation
\begin{align}
    \log p_\theta(x_0) = \log p(x_1) - \int_0^1 \nabla_x \cdot v_\theta(x, t)\, dt\;.
\end{align}
\subsection{Autoregressive Models}
All the generative models discussed so far learn to transform samples from a simple latent distribution into the target data space. Autoregressive models take a fundamentally different approach. Rather than learning such a transformation, they model the target density directly by factorizing it via the chain rule of probability
\begin{align}
    p_\theta(x) = \prod_{i=1}^{n} p_\theta(x_i | x_{<i})
\end{align}
where $n$ is the dimensionality. 
Each conditional $p_\theta(x_i | x_{<i})$ is parameterized by a one-dimensional flexible functional form whose parameters are outputs of a neural network that takes all preceding components $x_{<i}$ as conditional inputs. The parametrization should ensure fast training and inference time and enough flexibility. The concrete choice, however, must be adequate for the data at hand. Large language models, for instance, operate over a discrete vocabulary and model each conditional as a categorical distribution over tokens. Physics data, in contrast, are typically continuous and require continuous parametrizations such as Gaussian mixture models~\cite{Butter:2023fov} or binned histograms with sufficiently fine resolution~\cite{Finke:2023veq,Birk:2024knn}. 
The sampling proceeds sequentially. We start by drawing $x_1 \sim p_\theta(x_1)$, then $x_2 \sim p_\theta(x_2|x_1)$ and so on until the full data point is constructed. Here, each $x_i$ may represent a single scalar component or a higher-dimensional object such as the four-momentum of a particle. This autoregressive structure provides exact and tractable likelihoods by construction by simply evaluating and multiplying the conditionals. However, the sequential nature of both sampling and density evaluation can become a bottleneck, as the computational cost scales linearly with the number of autoregressive steps.   \\\\
In practice, the choice of generative network depends on the application. For sample generation, diffusion models and in particular CFMs have emerged as the current state of the art in terms of modeling performance. However, since inference requires solving an SDE or ODE, their sampling time can be considerably longer than for single-pass models. For density estimation tasks, INNs remain the preferred choice, as their invertible structure provides direct and fast access to the learned density without requiring numerical integration. Autoregressive models also provide competitive performance and exact likelihoods by construction, but are less widely adopted in physics applications, possibly due to the need for explicit parametrization of each conditional and the sequential nature of sampling. GANs and VAEs are less commonly used for state-of-the-art generative tasks today, though VAEs are sometimes combined with other models, such as diffusion models, to enable dimensionality compression of the data space before generation.
\section{Use Cases of Generative Networks}
\label{sec:use_cases}

In the following, we organize applications by their primary use of the generative model, as a fast sampler or as a density estimator, although in practice many applications exploit both capabilities. We do not attempt to exhaustively cover all generative machine learning applications in fundamental physics, but rather highlight exemplary use cases to illustrate the breadth of the field.

\subsection{Fast Simulation}
Modern precision measurements in physics are reliant on precise simulations encoding our theoretical understanding. In particle physics, there is a well-established chain of different Monte Carlo simulators, each one responsible for mimicking a different aspect of a real-life collision at e.g. the Large Hadron Collider (LHC). 
First, event generators simulate the initial, hard scattering process of the colliding particles.
Then, parton shower simulators and hadronization model how the remnants of the hard scattering propagate.
This includes jet formation and initial decays.
Finally, detector simulation models how the generated particles interact with the trackers, calorimeters, and muon systems of a detector. 
These simulators come with a high computational cost that is predicted to outgrow the available CPU resources at the high-luminosity LHC. A promising approach to accelerate the generation is to augment the simulation with fast generated samples from a generative network. 
ML-based data augmentation has been implemented at every point in the simulation chain. 

For event generation, a subset of works aims to directly replace existing matrix element generators with generative machine learning models~\cite{Butter:2019cae, Otten:2019hhl, Butter:2021csz, Choi:2021sku, Butter:2023ira}. 
Similarly, generative models have been used to accelerate (and improve) hadronization modeling~\cite{Ghosh:2022zdz, Chan:2023ume, Ilten:2022jfm, Bierlich:2023zzd}.

However, the arguably most established use-case is detector simulations, motivated by the comparatively high cost of modeling particle-matter interaction. 
From an ML-perspective, detector simulations are challenging to model given their sparse and high-dimensional nature. 
Recently, a community challenge~\cite{Krause:2024avx} designed to foster research in this area showed impressive results for a variety of different generative networks in terms of accuracy and speedup compared to classical simulators. 
Generative detector simulations have been explored with various network architectures, spanning across Generative Adversarial networks~\cite{Paganini:2017hrr, Paganini:2017dwg, Vallecorsa:2019ked, Chekalina:2018hxi, ATLAS:2018wpe, Carminati:2018khv, Musella:2018rdi, Erdmann:2018kuh, Erdmann:2018jxd, Belayneh:2019vyx, Buhmann:2020pmy, Maevskiy:2020ank, Khattak:2021ndw, Buhmann:2021caf, ATLAS:2022jhk, Hashemi:2023ruu, Diefenbacher:2023prl, FaucciGiannelli:2023fow, Scham:2023cwn, Simsek:2024zhj}, 
Variational Autoencoders~\cite{Hoque:2023zjt, Liu:2024kvv, Smith:2024lxz}, 
Normalizing Flows~\cite{Krause:2021ilc, Krause:2021wez, Diefenbacher:2023vsw, Ernst:2023qvn, Schnake:2024mip, Buss:2024orz} 
as well as Diffusion Models and Conditional Flow Matching~\cite{Mikuni:2022xry, Buhmann:2023bwk, Acosta:2023zik, Imani:2023blb, Amram:2023onf, Mikuni:2023tqg, Buhmann:2023kdg, Kobylianskii:2024ijw, Favaro:2024rle, Giroux:2025mit, Buss:2025cyw, Buss:2025kiu}. \\
Rather than replacing individual steps of the simulation chain with dedicated surrogates, an alternative approach targets the particle-level (or detector-level) data space directly. Here, generative models learn to sample final-state particle distributions without explicitly constructing parton-level events or modeling the subsequent shower and hadronization steps. Jet constituents have been generated directly as particle clouds using, again, a variety of architectures, including GANs~\cite{Kansal:2021cqp, Buhmann:2023pmh}, normalizing flows~\cite{Kach:2022uzq, Kach:2022qnf}, diffusion models~\cite{Leigh:2023toe, Leigh:2023zle, Mikuni:2023dvk}, flow matching~\cite{Buhmann:2023zgc}, and autoregressive transformers~\cite{Finke:2023veq}. 
Beyond constituent-level information, generative models can even target the reconstructed event representation directly, generating jet four-momenta and other analysis-level observables without modeling any intermediate step of the simulation chain~\cite{Butter:2019cae, Hashemi:2019fkn,Butter:2021csz, Butter:2023ira}. \\
In cosmology, a structurally analogous simulation chain exists. Starting from the nearly uniform matter distribution of the early universe, N-body simulations evolve the initial conditions forward in time under gravity. This produces the present-day large-scale structure of our universe. To connect these predictions to actual observations, additional modeling is required. To this end, dark matter halos must be populated with galaxies through semi-analytic prescriptions or full hydrodynamical simulations that model gas cooling, star formation, and feedback. Finally, observational effects such as gravitational lensing, survey geometry, and instrumental noise must be applied, analogous to detector simulations. As in particle physics, this chain is computationally expensive, and generative surrogates have been developed at multiple stages. \\
GANs, diffusion models and conditional flow matching models have been used to emulate three-dimensional dark matter density fields from N-body simulations, either directly~\cite{Rodriguez:2018mjb,Troster:2019mys,Mustafa:2019cos, Perraudin:2019bxl, Perraudin:2020gig}, by modeling survey observables~\cite{Cuesta-Lazaro:2023zuk,Horowitz:2025rke, Pandey:2024otp,Nguyen:2024ndo}. Flow matching models have also been used to learn compressed representations of projected dark matter density fields for reconstruction and parameter inference~\cite{Kannan:2025cem}.
\subsection{Density Estimation}
There are many physics applications where using a generative network to explicitly estimate an unknown density can make sense.
\subsubsection*{Anomaly Detection}
Anomaly detection was one of the earliest applications of generative models in particle physics, with (variational) autoencoders used to learn the density of known Standard Model backgrounds~\cite{Heimel:2018mkt, Farina:2018fyg}. Events poorly reconstructed by the decoder, or equivalently, events with low likelihood under the learned model, are flagged as anomalous. More recent approaches use the explicit density access or samples from the background distribution directly provided by the generative network to threshold on the learned density~\cite{Nachman:2020lpy,Hallin:2021wme,Belis:2023mqs}. The same procedures can also be used in astrophysical data such as observations made with the Gaia telescope to find stellar streams~\cite{Shih:2021kbt,Shih:2023jfv,Sengupta:2024ezl,Hallin:2025wyc}. In cosmology, discrepancies between theoretical models and observational data have been identified using normalizing flows to flag out of distribution samples~\cite{Dai:2023lcb,Diao:2025szg, Akhmetzhanova:2025kdb}.  
A detailed treatment of modern anomaly detection algorithms is provided in a companion VERaIPHY contribution~\cite{phystat-ad}.
 \\
\subsubsection*{(Un)folding}
Problems like unfolding, where the true particle-level distribution is inferred from smeared detector-level measurements, can be naturally framed as a density estimation problem. Here, generative networks are used to learn the conditional density $p(x_\text{part} | x_\text{reco})$, enabling probabilistic unfolding that captures the full posterior rather than a single point estimate~\cite{Datta:2018mwd,Bellagente:2019uyp, Bellagente:2020piv, Shmakov:2023kjj,Shmakov:2024gkd,Huetsch:2024quz,Butter:2024vbx,Butter:2025mek, Petitjean:2025tgk}. Similarly, given parton- or particle level distributions, generative networks also allow skipping the detector entirely and modeling reconstructed sets of objects directly~\cite{DiBello:2022lns,Kobylianskii:2024sup,Dreyer:2024bhs, Dreyer:2025zhp}. 
In addition, generative networks trained on simulation have been used combined with physical forward models to sample from the posterior distribution of cosmological inverse problems, such as the reconstruction of source galaxies from gravitational lensing observations~\cite{Adam:2022zji}, dark matter mass mapping from weak lensing data~\cite{Remy:2022ixn}, gravitational lensing potential from distorted Cosmic Microwave Background observations~\cite{Floss:2024gqk}, and the separation of astrophysical sources from multi-view observations~\cite{Wagner-Carena:2025yhe}. More details on inferring distributions can be found in a companion VERaIPHY contribution~\cite{phystat-sbi}.
\subsubsection*{Simulation-Based Inference}
More broadly, likelihood estimation is central to many inference tasks in physics. Generative models can in principle approximate any likelihood $p(x|\phi)$ as a function of theoretical parameters $\phi$ or $p(\phi|x)$ as a function of observation $x$, enabling downstream statistical analysis. Recent approaches in particle physics directly target the likelihood with respect to parameters~\cite{Butter:2022vkj,Heimel:2023mvw}, connecting generative modeling to simulation-based inference (SBI). In cosmology, neural SBI through learning either the likelihood or the posterior has become a well-established tool in the community. Most applications try to infer cosmological parameters from astrophysical observations~\cite{Alsing:2019xrx,Jeffrey:2020xve, Hahn:2022zxa, Mudur:2024fkh, Legin:2023jxc, Dai:2023lcb}. For details, see also the companion VERaIPHY contribution on SBI with machine learning~\cite{phystat-sbi}.
\subsubsection*{Performance Limit Quantification}
Determining fundamental performance limits is another application of density estimation. By training generative models with tractable likelihoods on signal and background samples, one can compute the Neyman-Pearson optimal likelihood ratio and compare it against existing classifiers. Recent work on jet tagging has used this approach to assess how close state-of-the-art taggers are to the theoretical optimum~\cite{Geuskens:2024tfo,Pang:2025lbs}.
\subsubsection*{Neural Importance Sampling}
Another application is neural importance sampling, which numerically solves Monte Carlo integrals with generative networks learning a proposal distribution to reduce variance and speed up convergence. This is used e.g. to speed up complicated cross section calculations for LHC simulations~\cite{Gao:2020vdv, Gao:2020zvv,Bothmann:2020ywa,Heimel:2022wyj, Heimel:2023ngj, Heimel:2024wph, Deutschmann:2024lml,Bothmann:2025lwg, Janssen:2025zke, DeCrescenzo:2026tsp}. The same idea has been applied in lattice field theory, where high-dimensional path-integral expectation values are conventionally estimated by Markov chain Monte Carlo. Normalizing flows are trained as proposal distributions approximating the Boltzmann weights~\cite{Albergo:2019eim, Albergo:2022qfi,Gerdes:2022eve, Abbott:2023thq, Abbott:2024kfc, Abbott:2024mix, Abbott:2025kvi, Bonanno:2025pdp} as well as diffusion models~\cite{Wang:2023exq,Zhu:2025pmw}.
\subsubsection*{Super-Resolution}
Lastly, generative networks have been used to perform super-resolution, i.e.~enhancing the resolution of a measurement or simulations beyond what was originally recorded or computed, by learning the conditional probability of high-resolution output given low-resolution input $p(x_\text{fine}| x_\text{coarse})$~\cite{Erdmann:2023ngr,Pang:2023wfx}. Conceptually, this is closely related to unfolding, with both tasks learning to invert a resolution-degrading process.
\section{Validation}
\label{sec:validation}
To validate the accuracy of a generative network, we need to ensure that it learned to reproduce the underlying distribution well and does not mismodel any regions of the data space.
Mismodeling refers to a systematic discrepancy between the learned generative density and the true underlying data distribution. Such discrepancies can arise from limited model expressivity, limited training data, imperfect optimization, or numerical precision constraints.
The ability of a generative model to accurately represent complex distributions depends on its architecture. For example, by construction flow-based models apply smooth and differentiable transformations to a base distribution. As a result, they struggle to represent distributions with sharp discontinuities or disjoint support. Thus, step-like structures are typically smoothed out, leading to small interpolations in the modeled density near sharp edges. Similarly, multi-modal distributions may be represented as approximately continuous regions with residual density in otherwise empty areas of teh data space. Multi-modal distributions can also be challenging for architectures like GANs, which are prone to mode collapse,~i.e. instead of mapping out the whole distribution, they learn to sample from a single mode.
Besides architectural constraints, mismodeling can also be caused by finite computational resources, which can limit both the expressivity and convergence of a generative network. Reducing the number of model parameters or training time often prevents the model from reaching an accurate approximation of the data distribution. Furthermore, optimization itself can lead to mismodeling, in cases such as vanishing gradients, even in expressive architectures. \\
%
Importantly, the source of mismodeling is often unknown a priori, and no single validation tool is universally optimal for detecting all types of discrepancies. 
In some cases, validation can exploit direct density access, for example by knowing a closed form expression of the underlying density of the true test data $\pdata(x)$. Here, common one-dimensional goodness-of-fit procedures like one-sample Kolmogorov-Smirnov tests can be used to validate whether the generated samples $x \sim p_\theta (x) \equiv \pgen(x)$ follow $\pdata(x)$. 
In other cases, the model provides a tractable likelihood, so that validation can proceed by evaluating $\pgen(x)$ directly on test data. The test set negative log-likelihood provides a global quality measure. 
However, in most practical settings we only have samples from both distributions. Moreover, even for models with tractable likelihoods, sample-based tests remain useful as they probe different failure modes.\\
In two-sample tests, we compare samples drawn from $\pgen(x)$ and $ \pdata(x)$ by evaluating their similarity through a given metric. In practice, the two distributions will never match exactly, and with sufficient generated samples any statistical test will eventually detect remaining discrepancies. The practical goal of validation is therefore to quantify the magnitude and location of remaining discrepancies rather than to test a binary hypotheses. Ideally, the chosen metric should also provide interpretable diagnostics when the distributions differ. In the following, we always assume that the reference samples $x\sim \pdata(x)$ are drawn from a held-out test set not used during the generative model training. 
Different metrics and tests are sensitive to different failure modes, some excel at detecting global distributional shifts, while others are more powerful for localized discrepancies or subtle correlation mismodeling. This motivates the use of a multi-pronged validation strategy, combining complementary approaches:
\begin{itemize}
    \item \textbf{Physics-informed checks} for domain-specific interpretability, providing the most direct and intuitive diagnostics,
    \item \textbf{Global metrics} for quick, global and high-dimensional comparisons of distributional similarity,
    \item \textbf{Classifier-based tests} for phase-space-resolved diagnostics that can localize mismodeled regions in high dimensions.
\end{itemize}
%
These categories are not mutually exclusive but rather represent complementary building blocks that can be combined. For instance, global metrics can be evaluated on classifier weights rather than raw observables, histogram-based chi-squared tests can be applied to the weight distribution, and a data space dependent classifier can be summarized into a single global test statistic. The categorization above reflects the primary diagnostic mode of each approach, but in practice a robust validation strategy will often combine elements from multiple categories. We discuss each category in detail in the following subsections. \\\\
Before discussing specific metrics, it is important to distinguish between the formulation of a test statistic and its calibration used to assign it a p-value. A test statistic quantifies distributional similarity according to a specific criterion, while calibration, e.g. through permutation tests, asymptotic approximations, or bootstrapping, converts the observed value of the test statistic into a statistical statement about the null hypothesis ($\pgen(x) = \pdata(x)$). Permutation tests, for instance, are a general and broadly applicable calibration method. Here, the labels distinguishing the two samples are randomly shuffled many times and the test statistic is recomputed for each permutation, yielding an empirical null distribution against which the observed value can be compared. They can be used with any test statistic to obtain valid p-values without relying on distributional assumptions.
\subsection{Physics-Informed Validation}
The most immediate and interpretable validation of a generative model exploits domain knowledge. Visually inspecting one-dimensional marginal distributions and two-dimensional correlations between physically relevant observables is often the first diagnostic step. A standard approach involves binning the modeled and true samples into histograms and evaluating their bin-wise agreement through a $\chi^2$-test. In practice, the resulting $\chi^2$ value depends on the choice of binning and the sample sizes of both generated and reference data, making it difficult to interpret as an absolute quality measure. It is most useful as a comparative metric at fixed samples sizes, for instance when ranking different generative models against the same reference data. 
Beyond marginals, one can verify that generated samples respect known conservation laws (e.g. energy-momentum conservation), reproduce expected cross-section ratios, or match analytic predictions for specific moments of the distribution. Such physics-motivated checks are often the most intuitive diagnostics for domain experts and can catch issues that purely statistical tests might miss, for instance a generative model that statistically matches the data on average but violates a physical constraint.
However, as the dimensionality increases, visual inspection and the evaluation of individual marginals become insufficient. Discrepancies may be confined to small regions of the high-dimensional data space or may involve non-trivial correlations between variables that are not captured by any single marginal or pair of observables. This motivates the use of more formal high-dimensional two-sample tests.
\subsection{Global Metrics}
Several global metrics have been developed to summarize high-dimensional distributional similarity. These can be broadly divided into two categories based on their statistical properties.
The Fréchet Inception Distance (FID)~\cite{FID} and its physics analog, the Fréchet Physics Distance (FPD)~\cite{Kansal:2022spb}, compare distributions by modeling them as Gaussians in a feature space extracted by a neural network and computing their Wasserstein-2 distance. While computationally efficient, this Gaussian assumption limits their statistical interpretability. The resulting distance values do not directly yield p-values or confidence intervals, and the metric can be insensitive to non-Gaussian features of the distributions.
A more principled framework is provided by the Maximum Mean Discrepancy (MMD)~\cite{Gretton:2008ker}
\begin{align}
    \text{MMD}^2(\pdata,\pgen) = \langle k(x,x^\prime) \rangle_{x,x^\prime \sim \pdata} 
    + \langle k(y,y^\prime)\rangle_{y,y^\prime \sim \pgen} - 2 \langle k(x,y)\rangle_{x \sim \pdata, y\sim \pgen} \;,
\end{align}
where $k$ is a positive definite kernel. This is a kernel-based two-sample test statistic that can measure the distance between distributions by comparing all moments simultaneously.
The Kernel Inception Distance (KID)~\cite{KIP} and its physics analog, the Kernel Physics Distance (KPD)~\cite{Kansal:2022spb}, are concrete implementations of the MMD in learned or physics-motivated feature spaces. Crucially, MMD-based metrics admit unbiased estimators with known asymptotic distributions, enabling rigorous hypothesis testing with well-defined p-values and confidence intervals~\cite{Gretton:2008ker}. This makes KID/KPD considerably more statistically interpretable than their Fréchet-based counterparts.
Another well-known test statistic is the energy distance~\cite{szekely2004testing}, which is related to the MMD with a specific kernel choice. Sliced versions of the Wasserstein and Kolmogorov-Smirnov distances have also been proposed and studied for high-dimensional generative model validation~\cite{Coccaro:2023vtb, Grossi:2024axb}.
The key limiting factor common to all global metrics is their lack of local sensitivity. They summarize distributional agreement into a single number, which makes them interpretable as global quality measures but potentially insensitive to highly localized discrepancies. In principle, one can evaluate these metrics in different regions of the data space, but such regions must be defined by cutting on selected directions, making the procedure less agnostic. A systematic comparison of the sensitivity and computational trade-offs of various global two-sample tests for generative model validation in high-dimensional settings can be found in Refs.~\cite{Coccaro:2023vtb, Grossi:2024axb}. Nevertheless, in some downstream tasks global metrics remain useful, as illustrated in Sec.~\ref{sec:amplification}.
\subsection{Local Metrics}
\begin{figure}[b]
    \centering
    \includegraphics[width=0.5\linewidth]{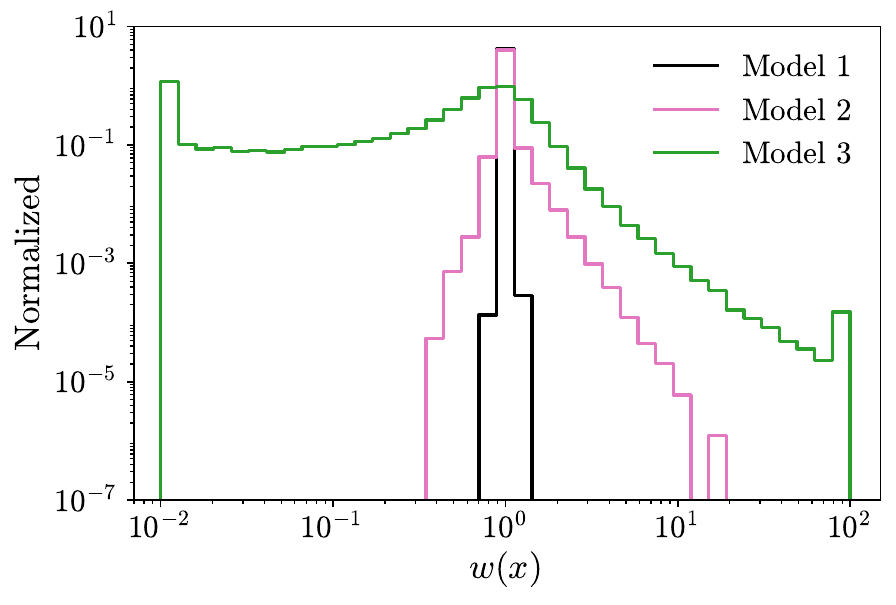}
    \caption{Distributions of classifier weights illustrating different levels of generative model quality. Each line represents a distribution sampled from a different hypothetical generative model.}
    \label{fig:classifier_weights}
\end{figure}
A more informative alternative and increasingly popular approach is to train a classifier to distinguish samples generated by the model,
$x \sim \pgen(x)$ from true data samples $x \sim \pdata(x)$~\cite{Friedman:2003id, Lopez:2018cts,Grosso:2023scl,Das:2023ktd}. In this setup, the converged classifier output $C_\theta(x)$ is directly related to the ideal test statistic, i.e. the likelihood ratio, through
\begin{align}
    w(x) \equiv \frac{\pdata(x)}{\pgen(x)} \approx \frac{C_\theta(x)}{1-C_\theta(x)} \;,
\end{align}
where the approximation becomes exact in the limit of infinite training data and sufficient classifier capacity.
The likelihood ratio as a function of $x$ provides a powerful one-dimensional diagnostic. It compresses all the relevant multidimensional information into a single scalar variable while retaining optimal sensitivity in the sense of Neyman–Pearson. Provided the classifier is well-trained and has sufficient capacity, the weights $w(x)$ for $x\sim \pgen(x)$ reveal
the presence of localized regions that are systematically mismodeled. Large values of $w(x)$ indicate regions where the data density exceeds the model density, while small values signal regions where the model overshoots.
Moreover, it can also be related to the global quality of the generative model across the full distribution. The distribution of the weights itself, like its mean, width, and tail behavior, summarizes how well the model matches the data overall. A narrow peak around $w=1$ indicates good agreement, whereas broadening or heavy tails indicate systematic discrepancies, mode-dropping, or distortions in the learned distribution.
A good generative model should therefore yield weights $w(x)$ that are tightly clustered around unity, indicating that the modeled distribution matches the true data distribution in the bulk of data space. Any discrepancies should manifest only in the tails of the distribution, and these tails should decay rapidly, corresponding to rare and localized regions of mismodeling. Moreover, evaluating $w(x)$ on samples drawn from the true distribution, $x \sim \pdata(x)$ should yield a weight distribution that is close to the one obtained from generated samples $x \sim \pgen(x)$. 
To not only identify but also understand mismodeled data regions, sampled points $x \sim \pgen(x)$ that yield very low or very high weights can be analyzed further by applying a cut on the weight distribution and investigating their distribution in the data space.
In addition to the weight distribution, the classifier used to construct $w(x)$ provides a natural global performance metric, the area under the receiver operating characteristic curve (ROC) or the AUC score. The ROC curve is evaluated at thresholds $\tau \in [0,1]$ of the classifier output. At each value for $\tau$ the true positive rate (TPR) and false positive rate (FPR) are evaluated, resulting in a curve (FPR($\tau$), TPR($\tau$)) from (0,0) to (1,1). Its integral, the AUC, summarizes how well the classifier can discriminate between real and generated samples. An AUC value close to 1 signifies that the two distributions are easily distinguishable, revealing substantial mismodeling. Conversely, an AUC value near 0.5 indicates that the classifier cannot do better than random guessing, consistent with the case where the generative model reproduces the data distribution faithfully or the classifier has failed to learn the relevant differences. \\
Since the diagnostic power of this approach relies entirely on the quality of the trained classifier, it is important to verify that the classifier has learned the correct likelihood ratio. A simple closure test consists of reweighting the generated samples by $w(x)$ and verifying that the resulting distribution reproduces $\pdata(x)$ in one-dimensional marginals. While agreement in marginals is not sufficient to guarantee that the classifier has fully captured the likelihood ratio over the full high-dimensional data space, any visible disagreement immediately signals that the classifier is unreliable.\\
In Fig.~\ref{fig:classifier_weights}, we illustrate how the distribution of classifier weights can serve as a diagnostic for generative model quality by showing three hypothetical weight distributions. Model 1 is tightly peaked around unity, indicating that the generated distribution closely matches the true data distribution across the full data space. Model 2 shows extended tails toward large weights, corresponding to regions where the generative model underestimates the true density, i.e.~the data contains structure that the model fails to reproduce. Model 3 exhibits heavy tails in both directions, indicating that the model both overshoots the data distribution in some regions (small weights) and undershoots it in others (large weights), suggesting more severe and widespread mismodeling.  \\
This classifier formalism can be extended to construct a formal hypothesis test~\cite{Grosso:2023scl}, which has been applied to validate generative models in high-dimensional settings~\cite{Grossi:2025pmm, Cappelli:2025myc}, or to validate conditional distributions~\cite{Butter:2024vbx}. Here, the classifier can learn the likelihood ratio of the modeled and true conditional probability through the likelihood ratio of the joint distributions
\begin{align}
    w(x|c) = \frac{\pdata(x|c)}{\pgen(x|c)} = \frac{\pdata(x,c)}{\pgen(x,c)}
\end{align}
where $c$ is a condition of arbitrary dimensionality and $\pgen(c) = \pdata(c)$. 
Note that while this formulation provides a useful diagnostic for conditional distributions, the Neyman-Pearson optimality of the unconditional case does not directly extend to composite hypotheses parameterized by $c$.
For applications such as SBI, an additional question is whether the learned conditional, i.e. the posterior distribution, is well-calibrated. For example, if a model predicts a 68\% credible interval for a parameter $\phi$ given an observation $x$, the true value of $\phi$ should fall within that interval 68\% of the time. Coverage tests check this by drawing many pairs of $(\phi, x)$ from the joint distribution and verifying that the stated credible regions contain the true parameter at the expected rate~\cite{Lemos:2023sba}.
\section{Uncertainties}
\label{sec:Uncertainties}

In this section, we briefly cover uncertainty estimation for generative models. 
Since the topic of uncertainty quantification in ML is already covered in a dedicated work in this series~\cite{Haussmann:2026gbi}, we refer to it for an in-depth exploration of uncertainties. 
There are various sources of uncertainty when estimating densities with generative networks. First of all, the statistics of the training data are limited. We therefore expect a statistical uncertainty associated with the network output. 
Secondly, there is a level of stochasticity concerning the network training, as, e.g., optimization is generally achieved by stochastic gradient descent, and networks are initialized randomly.
Lastly, besides limited statistics, there are also inherent uncertainties in the training data or in its representation. 
If a generative network is trained to estimate the density of, e.g., the detector response of showers in the electromagnetic calorimeter, the uncertainties of the MC simulations will translate to uncertainties in the generative model. 
For example, if the training data is continuous but represented by discretized or voxelized detector hits for the purpose of training the generative model, then the generative network can only estimate the discretized data density, which presents an inherent deviation from the continuous training data. 
In this section, however, we will focus on the statistical and systematic uncertainties of the network. Therefore, we assume that the training data itself, as well as its representation, is a perfect description of nature or can be calibrated independently from the generative model. 

One further notable distinction is the difference between aggregate uncertainties and per-sample uncertainties. 
Aggregate uncertainties apply to sets of data. 
A common example in physics is bin uncertainties in a histogram; here, the uncertainty of each point contained in a bin is not necessarily known.
Sample uncertainties apply to each data point. 
These uncertainties have the advantage that they can be used to perform error propagation in downstream tasks. 
The type of uncertainty quantification that is preferable is highly dependent on the task. 
For example, fast simulation, mirroring classical simulation, is only interested in aggregate uncertainties, as per-sample uncertainties are largely meaningless.
On the other hand, single-shot text generation tasks have no use for overall uncertainties and require a confidence score for each text sample instead. 

In the following, we will briefly discuss approaches to quantify the uncertainties of generative models that are used in particle physics. 

\subsection{Ensembles}

Ensembling refers to the practice of training multiple versions of networks with identical architecture on either the same training data or on variations of the training data.
In the generative case, each network is then used to generate samples, which are aggregated across the ensemble. 
Ensembling can only provide aggregate uncertainties. 
To this end, the samples from each network in the ensemble are turned into an aggregate quantity (such as histogram bin counts), and then the variance of these quantities is calculated.

The types of uncertainty encoded by ensembling depend on what parameters are varied in the ensembling process. 
If the same network architecture is repeatedly trained using the same hyperparameters but with different random initializations, then the ensemble over these networks will only encode the uncertainty introduced by the initialization.
Bootstrapping the training data for each ensemble network will add the statistical uncertainty of the training data into the ensemble uncertainty. 
Notably, uncertainties introduced by choosing an imperfect network architecture cannot be reliably captured by ensembling, as each training in the ensemble will be affected similarly. 

\subsection{Bayesian Neural Networks}
An alternative approach for incorporating statistical uncertainties into network training is Bayesian neural networks (BNNs).   
BNNs replace each (or most) trainable network parameter with a Gaussian distribution encoded by a mean and a width. 
Each time the network is evaluated during training, a value is sampled from these distributions using the reparameterization trick to ensure backpropagation.
Training a BNN requires an additional regularization loss term to prevent the Gaussians from collapsing into delta distributions. 

Further, the BNNs make the assumption that the loss of the underlying model is a likelihood loss. 
This can limit the application of BNNs for certain generative models.
When generating samples from a generative BNN, each network parameter is again sampled from the Gaussian distribution, but this sampling is kept constant for the generation of a set.
This sampling process is repeated with different instantiations of the network weights, each time producing a new set.
Calculating the variance over these sets, similar to what is done in ensembling, produces aggregate uncertainty predictions.
For generative architectures like a normalizing flow or conditional flow matching, which produce a likelihood prediction for a given data point, a BNN can obtain an uncertainty estimate on the predicted likelihood of a data point. 
With this, Bayesian flow models can also be used to obtain per-sample uncertainties~\cite{Butter:2021csz}.

BNNs have been demonstrated to encode both statistical uncertainties of the training data~\cite{Bollweg:2019skg} and the systematic modeling uncertainties in their uncertainty predictions~\cite{Butter:2021csz}


\subsection{Calibration}

In order for any uncertainty estimation to be useful, one needs to ensure that the estimation is correctly calibrated. 
In the context of uncertainty estimation, this calibration process mostly refers to ensuring correct coverage.
For example, if an uncertainty quantification method returns a 90\% confidence interval, then the true value should be contained within the interval 90\% of the time. 
Any less, and the uncertainty interval is underestimated, any more and the interval is too large. 
For classification tasks, several approaches to this calibration have been explored.
With generative models, however, one faces the added complication that the calibration objective is not clearly defined and is dependent on the type of uncertainty that is estimated.

For aggregate uncertainties, such as per-bin uncertainties, it is possible to define the coverage of those bin uncertainties and use these for calibration. 
Here, one generates multiple, separate data sets from the same model, histograms each, and then calibrates the per-bin uncertainties in such a way that they contain the separate bin counts the appropriate number of times~\cite{Bieringer:2024nbc}. 
For example, for a 1$\sigma$ or $68\%$ uncertainty interval, one needs to ensure the repeated bin counts lie inside the interval $68\%$ of the time. 
This approach to calibration, however, only covers uncertainty sources that repeated sample generation is sensitive to, meaning systematic issues in the model training cannot be captured. 

For per-sample uncertainties, calibration is even more complicated, as a generative model cannot be used to produce the "same" sample multiple times. 
This makes the definition of coverage itself inherently challenging.

An alternative approach to calibration uses multiple real validation data sets to determine the coverage of the prediction uncertainties. 
This has the advantage of being sensitive to all uncertainties of the generative model, but it requires a large data set to be used for this validation. 
As such, calibrating on validation data is a promising approach for benchmarking uncertainty quantification approaches.
However, in applications such as fast simulation, where the very motivation for using the generative model is the high cost of generating data, calibrating on a validation set is no longer viable.

\section{Amplification}
\label{sec:amplification}

In this section, we discuss the question of \textsl{How many events we can generate using a model before additional events contain no more new information compared to the original training dataset?}
This is a fundamental question that has arisen repeatedly at multiple conferences and workshops in machine learning and statistics. 
As such, we decide to dedicate a longer section to this topic, with the goal of providing a comprehensive overview. 

In Section \ref{sec:Uncertainties} we introduced several sources of uncertainties, split into systematic and statistical, that affect generative models, as well as ways to estimate these uncertainties. 
However, there is an additional uncertainty contribution: \textsl{Generalization Uncertainty}.
The generalization uncertainty describes how a given network interpolates between and extrapolates beyond the original training data. 
For example, when a classification network is trained on simulated data and is subsequently applied to measurement data, the generalization uncertainty describes how well the network generalizes from simulation to measurement. 
For generative networks, normal definitions of extrapolation are not directly applicable.
Here, one generally does not have a test set to which the network should be applied after training. 
Instead, the generative generalization uncertainty manifests when extrapolating in resolution space. 

A fundamental challenge in generative modeling is that we aim to model the true density $\pdata$ using the density of the generative network $\pgen$.
\begin{align*}
     \pdata(x) \;&:\; \text{True density} \\
     \Ddata^{\Ntrain} \sim \pdata(x)\;&:\; \text{Set with } \Ntrain \text{ points, drawn from }  \pdata(x) \\
     \pgen(x) \approx  \pdata(x)\;&:\; \text{Generative model trained on } \Ddata^{\Ntrain} \text{, approximating } \pdata(x)\\
     \Dgen^{\Ngen} \sim \pgen(x)\;&:\; \text{Set with } \Ngen \text{ points, drawn from }  \pgen(x) 
\end{align*}
However, at no point do we have access to $\pdata$. 
Instead, we need to approximate the true density using the training data $\Ddata^{\Ntrain} \sim \pdata$ drawn from it. 
Any training set $\Ddata^{\Ntrain}$ will have a finite resolution defined by the number of points $\Ntrain$ contained within. 
Drawing more samples from the generative model than there are in the training data can be seen as extrapolating this finite resolution. 
\begin{align*}
     \Ngen > \Ntrain &:\; \text{Extrapolation in resolution}
\end{align*}
The generative generalization uncertainty provides a measure of how close the generative model is to the true distribution and, therefore, describes how well a network performs this extrapolation in resolution space. 
Inextricably linked to this uncertainty is the question: \textsl{How many events can we generate using a model before additional events contain no more new information compared to the original training dataset?}
If the generative model can generate more meaningful samples than those contained in the training set of the model, we say that the model \textsc{amplifies} the dataset. 
Defining what counts as a meaningful sample is not trivial.
Instead, it is more convenient to define amplification through the agreement of the generative model with the true density it is trying to learn.
If the generated set $\Dgen^{\Ngen}$ is closer to $ \pdata(x)$ (for sufficiently large $\Ngen$) than the training set $\Ddata^{\Ntrain}$, then the generative model amplifies the data. 

Conceptually, amplification is a core assumption of many generative applications, especially fast simulation. 
Here, the goal is to train a generative model on a small training set and then generate the bulk of the needed simulation from the model.
This approach requires the model to amplify the training set.
Without this amplification, the training data would be a more precise source of simulation than any amount of generated data.

\subsection{Quantifying Amplification }

Quantifying if and how much a model amplifies a data set requires us to quantify the agreement between a generated set and the true, underlying distribution. 
Following~\cite{Bahl:2025ryd} we introduce the arbitrary metric $M[D, p(x)]$, which measures the agreement between a set $D$ and a density $p(x)$. 
\begin{align*}
    M[D^{n}_{q}, p(x)] &:\; \text{Metric} \\
    p(x), q(x) &:\; \text{Densities}\\
    D^{n}_{q} \sim q(x) &:\; \text{Set with size }n\text{ sampled from  }q(x)\\
\end{align*}
We focus on a set of specific metrics in this chapter, however, any of the metrics described in Section~\ref{sec:validation} can, in principle, be used here. 

The contributing factors to $M$ can be split into two conceptual parts:
\begin{align}
    M[D^{n}_{q}, p(x)] &= \sigma_\text{stat}(D^{n}_{q}, p(x)) + \sigma_\text{model}(D^{n}_{q}, p(x)) \notag \\
    &= \sigma_\text{stat}(n) + \sigma_\text{model}(q(x), p(x))
    \label{eq:uncertainty_components}
\end{align}
\begin{align*}
    \sigma_\text{stat} &:\; \text{Statistical uncertainty from finite }n \\
    \sigma_\text{model} &:\; \text{Generalization uncertainty from } p \neq q
\end{align*}
Where the first term is the statistical fluctuation that arises from the finite size of the set, which only depends on $n$.
The second term is the mismatch between the two densities $p(x)$ and $q(x)$, in cases where $p(x) \neq q(x)$. 

Using this, we can define the expected behavior for the metric between a set drawn from a generative network and the true density
\begin{align}
    M[\Dgen^{\Ngen}, \pgen(x)] &= \sigma_\text{stat, \gen}(\Ngen) + \sigma_\text{model, \gen}(\pdata, \pgen)
    \label{eq:uncertainty_components_gen}
\end{align}
However, by itself, the value of $M$ is hard to interpret and highly unintuitive. 
One way to understand it is by relating it to the value of $M$ one expects from a set drawn from the true density
\begin{align}
    M[\Ddata^{n}, \pdata(x)] &= \sigma_\text{stat, \data}(n) + \sigma_\text{model, \data}(\pdata, \pdata) \notag \\ 
    &= \sigma_\text{stat, \data}(n) \; ,
    \label{eq:uncertainty_components_true}
\end{align}
where the second step uses the fact that $\sigma_\text{model}(\pdata, \pdata) = 0$ by definition. 
For a reasonably trained generative network, it is safe to assume that the training data and generated data are similar enough that the behavior of the respective statistical components is nearly identical. 
At the same time, we can rename the generalization uncertainty of the model for brevity
\begin{align}
    \sigma_\text{stat, \gen} &\approx \sigma_\text{stat, \data} = \sigma_\text{stat} \notag \\
    \sigma_\text{model, \gen} &= \sigma_\text{model} \;.
    \label{eq:stat_uncertainty_relation}
\end{align}
To summarize
\begin{align}
    M[\Dgen^{\Ngen}, \pdata(x)] &= \sigma_\text{stat}(\Ngen) + \sigma_\text{model} \notag \\
    M[\Ddata^{n}, \pdata(x)] &= \sigma_\text{stat}(n)
    \label{eq:uncertainty_components_summary}
\end{align}
Using this, we can now define the equivalent size of a generative model $\Nequiv$.
Intuitively, $\Nequiv$ can be understood as the number of points one has to sample from the true distribution to reach the same statistical uncertainty as the generalization uncertainty of the generative network. 
\begin{align}
    \sigma_\text{stat}(n) \Big|_{n=\Nequiv} \stackrel{!}{=} \sigma_\text{model}
\end{align}
In order to isolate $\sigma_\text{model}$ from $\sigma_\text{stat}(\Ngen)$, one can leverage the fact that the statistical component is expected to approach $0$ as the number of generated samples increases.
\begin{align}
    \sigma_\text{stat}(n)&\Big|_{n\to \infty} \to 0 \notag \\
    \sigma_\text{model}&\Big|_{n\to \infty} = \sigma_\text{model} \notag \\
    M[\Dgen^{\Ngen}, \pdata(x)] &\Big|_{\Ngen \to \infty} = \sigma_\text{model}
\end{align}
This leads us to the canonical definition of $\Nequiv$ for a given model density $\pgen$ and metric $M$
\begin{align}
    M[\Ddata^{n}, \pdata(x)] \Big|_{n = \Nequiv} \stackrel{!}{=} M[\Dgen^{\Ngen}, \pdata(x)] \Big|_{\Ngen \to \infty} 
\end{align}
Using this, we can define cases in which amplification occurs, as well as the amplification factor $G$, 
\begin{align}
    \Nequiv \leq \Ntrain &: \text{no amplification} \\
    \Nequiv > \Ntrain &: \text{amplification} \\   
     G=\frac{\Nequiv}{\Ntrain} &:  \text{amplification factor} 
\end{align}
A network with an amplification factor of 5 can therefore extrapolate to a resolution equivalent to that of a dataset 5-times larger than the original training set of the network.

\subsection{Practical Estimation}

Several approaches for estimating generative amplification have been proposed.
The main difference between methods are what exact metric they employ in place of $M$. 
Each approach has different caveats and limitations, and no clear winner that works for all applications has been found. 
Therefore, it is important to closely consider which methods work best for a given problem when aiming to test the amplification behavior of a generative network.
The downsides of each method are covered in the respective segments. 

\paragraph{Quantile Amplification Measure}

Initial studies on the topic directly used the true underlying distributions to demonstrate and quantify amplification~\cite{Butter:2020qhk, Bieringer:2022cbs, Bieringer:2024nbc}.
One of the metrics used is the Quantile Mean Squared Error
\begin{align}
    M\left[ \Dgen^{\Ngen}, \pdata(x) \right] &= \sum_{i=0}^{n_\text{quant}} \left(F_i - \frac{1}{n_\text{quant}} \right)^2 \\
    F_{i} &= \sum_{x \in \Dgen^{\Ngen}}\mathbf{1}_{q_{\text{low}, i} < x \in > q_{\text{up}, i}}  
\end{align}
where $\mathbf{1}$ is the indicator function and $q_{\text{low}, i}$ and $q_{\text{up}, i}$ are the lower and upper boundaries of the $i$-th quantile, respectively. 
These quantile boundaries are derived using the true distribution and are defined such that integrating the density over a quantile yields $\frac{1}{n_\text{quant}}$. Subsequent studies improved on the MSE metric by modifying it to a $r^2$ based metric, however these approaches were still based on the true density quantiles.

Using the true density allows those initial examinations to demonstrate that generative amplification is possible, however it limits their application to synthetic toy datasets, where the true distribution is known. 
The true distribution can be modeled using a sufficiently large holdout dataset. 
In this case, the distribution defined by a large set takes the role of the true distribution. 
Assuming the holdout set is sufficiently large compared to $\Ntrain$ and the expected $n_\text{eqiv}$, this method leads to a close approximation of the amplification factor.

The downside of the standard quantile approach is that it requires knowledge about the true distribution.
Many applications of generative models in particle physics have neither tractable truth distributions nor viable ways of producing a large holdout set.
The most notable example is fast simulation, where generating a holdout set significantly larger than the expected number of samples to be drawn from the generative model would defeat the very goal of fast simulation. 
In such cases, estimating the amplification factor of a model can be all the more important. 
Especially in fast simulation, it can be vital to know how many points can be drawn from the network before no further insight can be gained.

\paragraph{Averaging Amplification Measure}

\begin{figure}[]
    \centering
    \includegraphics[width=0.6\linewidth]{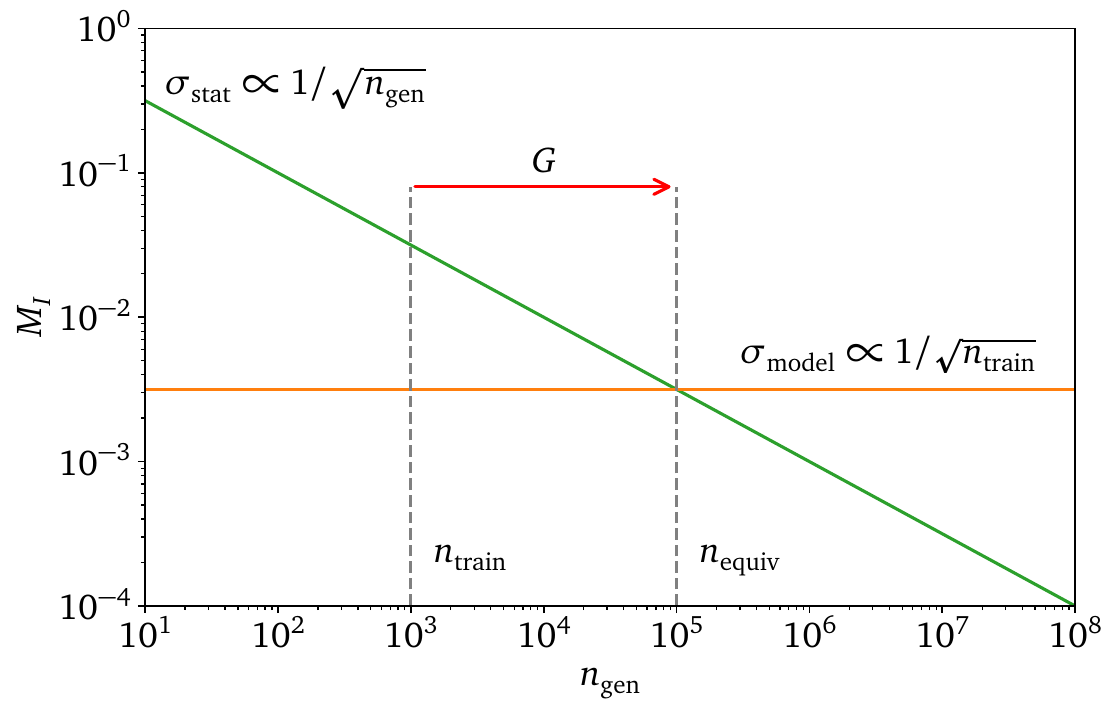}
    \caption{Schematic depiction of using the averaged amplification test to determine $\Nequiv$ and $G$. Taken from~\cite{Bahl:2025ryd}}
    \label{fig:MI_scheme}
\end{figure}

The averaging approach can be seen as an evolution of the quantile method.
As before, one splits the data into discrete sections $V_i$, with $i \in [1,n_\text{regions}]$. 
However, rather than comparing the counts in each region to the true values, one instead uses an uncertainty-aware generative network, such as a network ensemble or Bayesian neural network, to predict the variance of the fraction in each region. 
\begin{align}
  M_I \left[ D_{\gen}^{\Ngen}, \pdata\right]
  &= \langle \bar{I}^2 \rangle_{\theta} - \langle \bar{I} \rangle_{\theta}^2 \notag \\
\end{align}
Here $\bar{I}_{V_j, \theta_i}$ is the fraction of data points in the region $V_i$, drawn from the network with parameters $\theta_j$, where various $\theta_j$ parameter sets correspond to either different instantiations of the BNN or different networks in the ensemble. 
Explicitly, 
\begin{align}
    \bar{I}_{V_i, \theta_j}  =\frac{1}{\Ngen} \sum_{x \in D^{\Ngen}_{\gen, \theta_j}} \mathbf{1}_{x \in V_j} \;,
    \label{eq:bayes_integral}
\end{align}
with the indicator function $\mathbf{1}$.
Then, one leverages the fact that for an uncorrelated data set $\Ddata^{\Nequiv}$, a histogram representation of the regions would have a per-region Poisson uncertainty of 
\begin{align}
    \sigma_i = \sqrt{t_i} \quad \text{with} \quad t_i = \sum_{x \in  D} \mathbf{1}_{b_{\text{low}, i} < x \in > b_{\text{up}, i}}  \;.
\end{align}
Using the connection between the generative statistical uncertainty and the true-distribution statistical uncertainty postulated in Eq.~\ref{eq:stat_uncertainty_relation}, one can conclude that 
\begin{align}
    \sigma_\text{stat, \data}(n) \approx \sigma_\text{stat, \gen}(n) =\sigma_\text{stat}(n) \propto \frac{1}{\sqrt{n}} \;.
\end{align}
Since we expect $M_I \left[ \Dgen^{\Ngen}, \pdata \right]$ to be dominated by $\sigma_\text{stat}$ for $\Ngen \ll \Ntrain$, it is possible to derive the scaling of $\sigma_\text{stat}$ in relation to $n$ using the low-$n$ region.
This makes it possible to determine predicted values for $M_I \left[\Ddata^{n}, \pdata \right]$ without having to generate $\Ddata^{n}$.
From this, one can determine the crossover between $\sigma_{\text{stat}}(n)$ and $\sigma_{\text{model}}$, which occurs at $n=\Nequiv$.
Fig.~\ref{fig:MI_scheme} demonstrates this principle. 
The amplification factors derived from different regions $V_j$ can finally be combined by simple weighted averaging, resulting in an amplification factor for the whole distribution. 

The averaging method faces two downsides. 

(i) Determining $M_I$ requires an uncertainty-aware generative network, which is nontrivial. 
Further, the accuracy of the amplification measure is only as good as the accuracy of the predicted uncertainties. 
This was shown in Ref.~\cite{Bieringer:2024nbc}, where additional calibration on a test data set was needed to make the uncertainty predictions sufficiently accurate. 

(ii) The precise amplification value obtained for the full distribution will be sensitive to how the individual regions are defined. 
Therefore, the observable used for the splitting and the precise split need to be carefully considered and adapted to any given application. 

\paragraph{Differential Amplification Measure}

\begin{figure}[]
    \centering
    \includegraphics[width=0.6\linewidth]{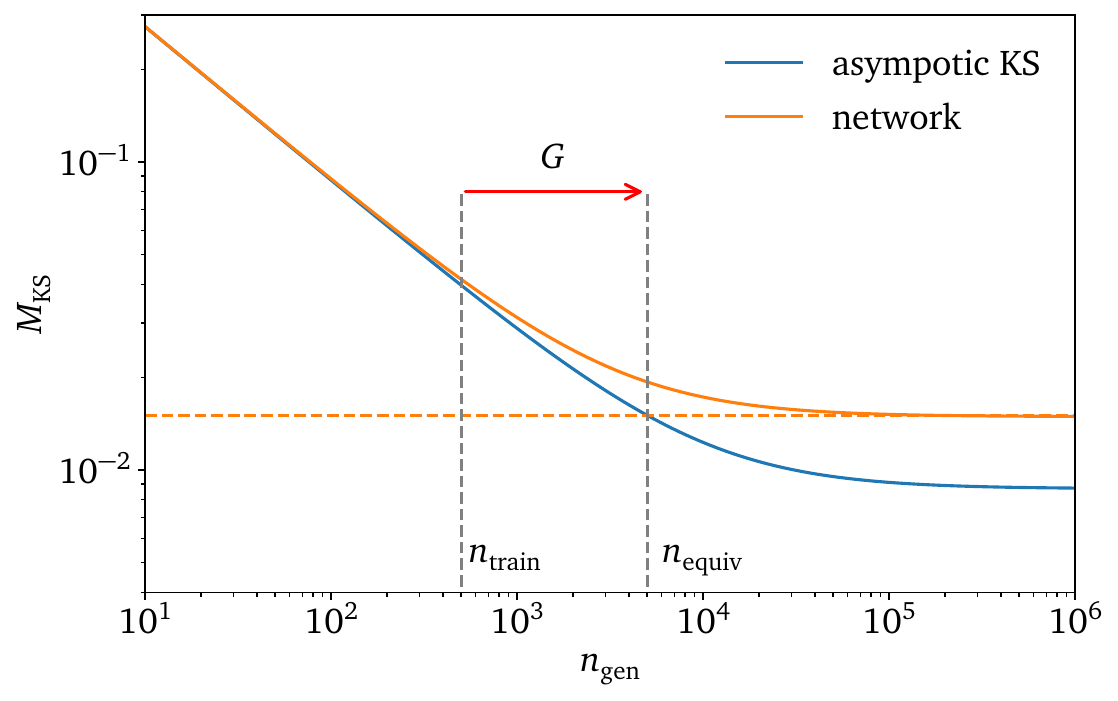}
    \caption{Schematic depiction of using the Kolmogorov Smirnov test to determine $\Nequiv$ and $G$. Taken from~\cite{Bahl:2025ryd}}
    \label{fig:KS_scheme}
\end{figure}

The second approach leverages the fact that a close approximation of the amplification factor can be achieved by replacing the metric $M[D, p]$ with $M^\prime[D, D^\prime]$, which instead operates on two datasets~\cite{Bahl:2025ryd}. 
Using this, one can define $\Nequiv$ using the training set $\Ddata^{\Ntrain}$
\begin{align}
    M^\prime\left[\Ddata^{n},\Ddata^{\Ntrain} \right] \bigg|_{n=\Nequiv} \stackrel{!}{=} 
    M^\prime\left[ \Dgen^{\Ngen}, \Ddata^{\Ntrain} \right]\bigg|_{\Ngen \to \infty}
    \; .
    \label{eq:gen_nequiv_KS}
\end{align}
The right-hand side of this equation can be evaluated directly using most standard divergences.
The left-hand side, however, is challenging as it requires a variable-sized comparison set $D_{\text{true}}^{n_{\text{eqiv}}}$, which potentially needs to be orders of magnitude larger than the training set. 
Similar to Eq.~\ref{eq:uncertainty_components_summary}, this can be broken down into:
\begin{align}
    M^\prime\left[ \Dgen^{\Ngen}, \Ddata^{\Ntrain} \right] &= \sigma_{\text{stat}}(\Ngen, \Ntrain) + \sigma_{\text{model}} \notag \\ 
    M^\prime\left[\Ddata^{n},\Ddata^{\Ntrain} \right] &= \sigma_{\text{stat}}(n, \Ntrain)
\end{align}
Importantly, since both datasets that are being compared have finite sizes, the statistical uncertainty now depends on both sizes. 
Combined, one can then once again equate
\begin{align}
    \sigma_\text{stat}(n, \Ntrain)\Big|_{n = \Nequiv} \stackrel{!}{=} \sigma_\text{stat}(\Ngen, \Ntrain)\Big|_{\Ngen \to \infty}  + \sigma_\text{model} \;.
    \label{eq:uncertainty_components_KS}
\end{align}
If the metric used has a predictable and analytically tractable behavior for the case where two data sets are drawn from the same distribution and only have statistical fluctuations, it is possible to predict the behavior $M^\prime\left[\Ddata^{n},\Ddata^{\Ntrain} \right]$ without the need to produce $\Ddata^{n}\big|_{n >> \Ntrain}$.
One such metric is the Kolmogorov Smirnov (KS) test, which has an analytical expression for the expectation value of $M_{\text{KS}}\left[\Ddata^{n},\Ddata^{\Ntrain} \right]$, given by
\begin{align}
    \left\langle M_{\text{KS}}\left[\Ddata^{n},\Ddata^{\Ntrain} \right] \right\rangle= \sigma_\text{stat}(n, \Ntrain) =\sqrt{\frac{\pi}{2}}\log 2\ \sqrt{\frac{n+\Ntrain}{n\;\Ntrain}}
    \label{eq:KS_analytic}
\end{align}
Through this, it is possible to directly evaluate Eq.~\ref{eq:gen_nequiv_KS} and thereby derive $\Nequiv$. 
Fig.~\ref{fig:KS_scheme} demonstrates this, showing the evaluated KS score for various values of $\Ngen$ as well as the analytical values calculated using Eq.~\ref{eq:KS_analytic}.
$\Nequiv$ can then be determined by finding the crossover between the analytical KS value and the KS value that the network reaches for $\Ngen \to \infty$.

The KS test approach has two main drawbacks, however.

(i) The KS test is only defined for one-dimensional distributions. 
This means it cannot be directly applied to most generative applications, which often feature high-dimensional data.
However, this drawback can be circumvented by training a classifier network to separate generated and true data.
This classifier will learn to represent the differences between the two datasets in a single observable, which can then be used as an input to the KS test. 
Since the differences the classifier learns are largely in line with the generalization uncertainty of the network, the classifier score is a close approximation of the full input space. 

(ii) As can be seen from Fig.~\ref{fig:KS_scheme} and Eq.~\ref{eq:KS_analytic}, the analytical value of $\sigma_\text{stat}(n, \Ntrain)$ quickly becomes dominated by $\Ntrain$. 
This causes $\sigma_\text{stat}$ to converge toward a fixed value for large $n$. 
Therefore, small variations in $\sigma_\text{stat}$ can correspond to large differences in $n$, which makes it challenging to precisely determine large amplification factors.

\subsection{Discussion}

Under a purely information-theoretical view, amplification seems highly unintuitive, as it seemingly produces information out of nowhere. 
In practice, however, amplification has repeatedly been demonstrated to occur for generative models, both for well understood toy distributions and for realistic physics data sets. 
Further, the closely related concept of \textsl{emulator superiority}~\cite{2025arXiv251023111K}, in which an emulator trained to solve partial differential equations can perform better than a high statistics reference despite being trained on low-statistics data, is discussed. 
This can be explained by viewing amplification as extrapolation in resolution space.
As a baseline, extrapolating with a neural network (i.e., applying the network to data outside of the domain it was trained on) would be expected to be ill-defined.
However, for a sufficiently smooth, regularized, and well-behaved network, it is still possible to provide reasonable results beyond its training regime. 
Amplification can be seen as a similar effect. 
One important caveat is, however, that a network will never be able to learn unique features that are not present in its training data.
In the amplification case specifically, this means that the network can learn the overall behavior of the training data and reproduce this well. However, any feature that is too small or too fine to be captured by the training data likewise will not be captured by the network. 
This was explored in more detail in Ref~\cite{Watts:2024caw}, where the Shannon entropy of histograms was used to derive a bound on possible amplification factors in the absence of inductive biases. 
Here, the authors found that 
\begin{align}
    2 \frac{\log(\Nequiv)}{\log(\Ntrain)} \equiv M_\text{eff} \;,
\end{align}
which relates to the amplification factor to a second factor $M$. This $M$ defines the relationship between the number of bins that result in a reasonable representation of a dataset $D^{n}$ in a histogram according to
\begin{align}
    n_{\text{bins}} = n^{\frac{1}{M}} \Leftrightarrow M = \frac{\log n}{\log n_{\text{bins}}}
\end{align}
Therefore, in the cases explored in~\cite{Watts:2024caw}, amplification is possible on an information-theoretical level.
However, it comes at the cost of reducing the viable number of bins and therefore reducing the resolution at which the  dataset can be trusted.
Specifically, the authors suggest $M_\text{eff} < 3$ as an upper bound, as higher values lead to significantly underbinned histograms. 

\begin{figure}[]
    \centering
    \includegraphics[width=0.6\linewidth]{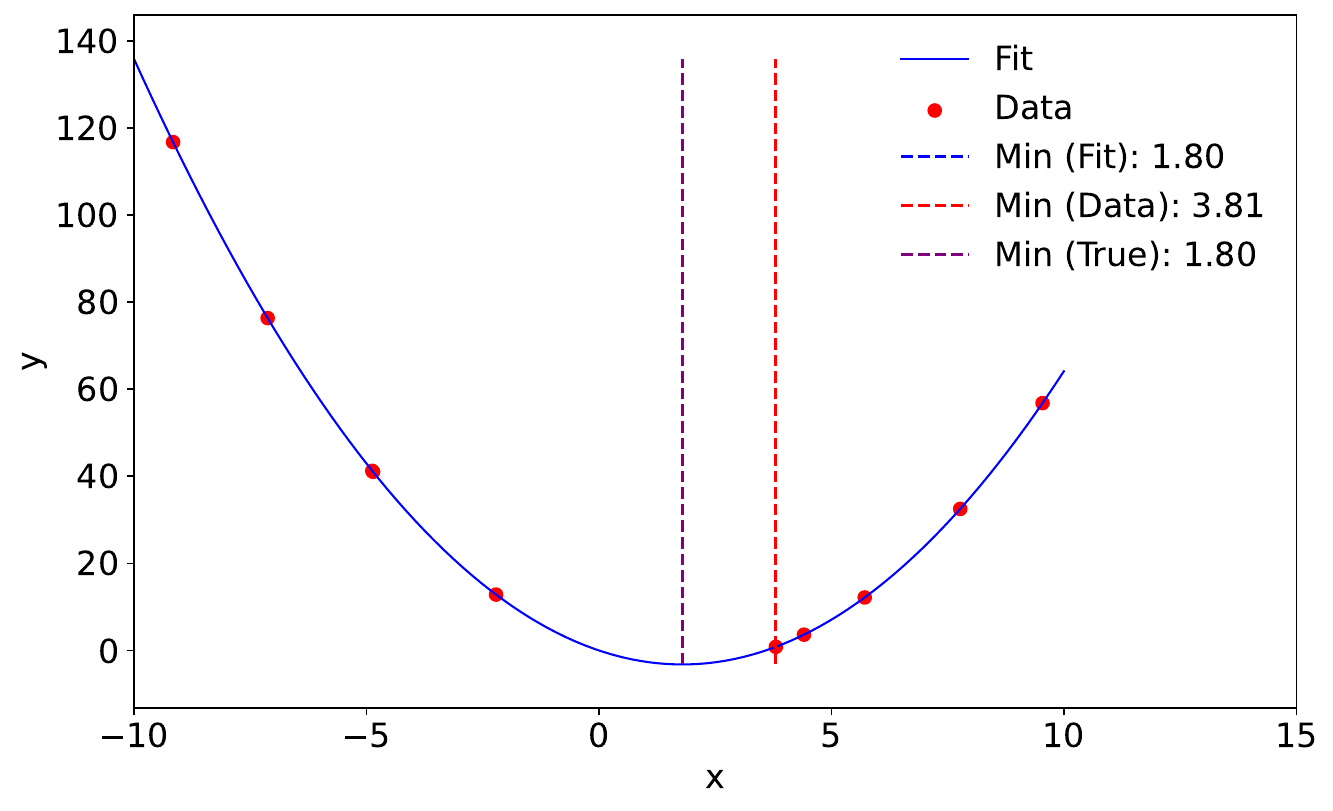}
    \caption{Comparison between determining the minimal value of a dataset drawn from a parabola and determining the minimum from a parabolic fit to that data. Used to illustrate the effects of inductive bias.}
    \label{fig:parabolic_fit}
\end{figure}

An important further consideration is that generative networks are not perfectly unbiased and do introduce inductive bias. 
To illustrate the effects of this inductive bias, we compare it to a parameter fit using the true functional form. 
As an example, we assume we have a small data set drawn from a parabolic function. 
We now compare how well we can use this dataset to determine the minimum of the distribution, and how well this can be achieved using a parabolic fit. 
For the dataset, we assume the minimum to be the lowest data point, and for the fit, we calculate the minimum from the resulting function. 
Fig.~\ref{fig:parabolic_fit} shows the result, where we can see that despite using the same 10 data points as its basis, the fit produces the minimum significantly better than the data alone. 
For generative models, this inductive bias is significantly weaker than that for a functional fit, as shown in Ref~\cite{Butter:2020qhk}.
Generative model biases are instead closer to smoothness constraints, which stem from the underlying neural networks. 
Nevertheless, this still helps the generative networks amplify the data if the smoothness biases hold true.

A further point to consider is limited network expressiveness. 
Our previous considerations assumed that the limiting factor for the agreement between the generative model and the true distribution is the training data.
This is a reasonable assumption in cases where the true distribution is simple enough for the network to fully capture it. 
In this case, amplification is possible even as we give the network an increasing amount of training data. 
Keeping with the parabola example from Fig.~\ref{fig:parabolic_fit}, even if one were to perform the fit on a number of data points approaching infinity, the two parameter parabolic fit would still be sufficient to describe the data. 
However, there are potential cases where the true distribution is too complex to be captured by a given network architecture. 
In this case, there is an upper limit to the $\Nequiv$ the network can reach, which cannot be increased by increasing the size of the training set. 
Nevertheless, this limit can still be exceeded by modifying the network architecture.
Disentangling these two contributions has not yet been investigated. 
Connected to this is the question of amplification for conditional generative models, which is likewise unexplored.

To summarize, generative amplification is viable in cases where one can be sure that the underlying smoothness assumptions of the network hold true and where one is confident that there are no important features in the true density that are not captured by the training data. 
Even in cases where this is not provided, it is possible to implement strong inductive biases into the generative network to push it toward a valid resolution extrapolation behavior. 
Therefore, generative models can reasonably be used in well understood and fully controlled settings, such as Monte Carlo simulation.
However, one cannot simply amplify experimental measurement data, which is not fully understood and likely holds unknown features. 
Furthermore, the exact mechanisms determining the amplification factor of a generative model and how one can improve this are still unclear. 

One final caveat is that any inductive bias the generative model introduces could also be leveraged in different ways. 
For example, if one first trained a generative model on the 10 data points in Fig.~\ref{fig:parabolic_fit} and then performed the parabolic fit on the generated data, one would not expect to see an improvement between the fit on the generated data and that on the original data. 
Similarly, if one had a perfect methodology for analyzing physics data or Monte Carlo simulations, such as perfect, tractable descriptions of the underlying distributions, generative amplification and generative simulation could become obsolete. 
However, this is not something we can achieve at this moment, and generative  simulation currently seems to be the most promising way to leverage this inductive bias to accelerate Monte Carlo simulations.

\section{Conclusion \& Outlook}
\label{sec:conclusion}
Generative models have the capacity to learn an underlying distribution from a set of data from a given sample and then produce new data points from the learned distribution.
Various types of generative models exist, which are fundamentally based on the concept of transforming an initial noise distribution into the learned data distribution. 
There is a wide spectrum of applications for generative models in fundamental physics, ranging from fast simulations across a wide range domains, to use in data analysis, to anomaly detection. 
The generative models in these applications have to contend with three statistical issues, which we explored in this paper. 

(1.) \textsl{Does the learned distribution represent the underlying distribution, and how can we validate this?} 
This question is directly linked to the quality of the generated samples and, therefore, the performance of the generative model. 
We have discussed several metrics that can be used to validate the agreement of a generative distribution with the target data.
A central challenge for these metrics is the often high dimensional nature of the data used in generative tasks.
Several classical metrics and divergences work well on one- or low-dimensional data but are challenging to efficiently generalize to high dimensions. 
As a result, several purpose-built generative metrics, such as the multiple Inception Distances or the Classifier Metric, first perform an embedding into a lower dimensional space before comparing generated and real data.
Especially in high energy physics, the classifier metric has become the gold standard for generative validation, as it simultaneously provides a clear-cut, single-number metric in the classifier accuracy (or AUC), while also providing more detailed insights based on the distribution of the classifier predictions. 

(2.) \textsl{How do we quantify the uncertainties of generative models, and how do we ensure they are well calibrated?} 
Fundamental physics research requires well quantified uncertainties to differentiate physics insight from random noise. 
Therefore, one will need a clear handle on generative uncertainties to successfully employ generative models in the long term. 
We discussed a range of generative uncertainty quantification approaches that have seen use, from ensembling over multiple model training to Bayesian generative models. 
A fundamental challenge of these approaches is that validating their calibration requires a large validation data set, which may not exist for applications like fast simulation.
As a result, generative uncertainties remain an important field of ongoing research.

(3.) \textsl{By how much does a generative model amplify its training set, and how can we optimize this?} 
The question of amplification is vital for any generative application that aims to sample new points beyond the training set. 
Especially fast simulation methods rely on the assumption that their amplification factors are greater than one. 
We explored several methods to define and quantify amplification. 
Earlier approaches struggled with similar issues to those faced by generative uncertainty quantification, as they required large comparison data sets to be useful.
More recent work has found ways to circumvent the need for such data sets; however, this comes with increased uncertainties regarding the amplification factors. 
This makes amplification quantification a field of active research with vital implications for the future use of generative models.

Generative networks are increasingly present across fundamental physics. 
As they mature from proof-of-concept studies to components of real analyses, it becomes essential that the physics community develops robust validation frameworks and understands how to interpret their shortcomings. 
While accuracy, uncertainty quantification, and amplification are often treated as separate challenges, they are deeply interconnected. 
Developing these tools will be critical as generative models take on greater responsibilities in physics workflows. 

Several important questions remain open:  
Can validation metrics be developed that scale to high dimensions without themselves relying on learned models? 
At what point does the systemic bias of a generative model outweigh the statistical gain from amplification, and can this threshold be determined before deployment? 
How do network imperfections propagate into downstream tasks, and how should network uncertainties be incorporated into those analyses? 
How robust are generative inference tools when applied to real data that may differ significantly from the simulations used for training? 
As generative models are deployed across different domains of fundamental physics, to what extent can validation frameworks be transferred between fields, and where do domain-specific solutions remain necessary? 
Answering these questions will be one of the central challenges in the coming years, as generative models become an integral part of how we do physics.
\section*{Acknowledgments}
This article is part of VERaiPHY (Validation \& Evaluation for Robust AI in PHYsics), a coordinated effort that unites researchers from fundamental physics, computer science, and statistics to discuss principled frameworks for assessing the reliability and scientific validity of modern ML methods. We would like to thank Gaia Grosso, Louis Lyons, and Ramon Winterhalder for comments on the initial draft, as well as our reviewers Theo Heimel, Yonathan Kahn, and Vinicius Mikuni. Lastly, we would also like to thank Carol Cuesta-Lazaro for providing an overview of relevant cosmological use cases. SPS acknowledges support from the DOE grant DOE-SC0010008.
GK acknowledges support by the Deutsche Forschungsgemeinschaft (DFG, German Research Foundation) under the German Excellence Initiative -- EXC 2121  Quantum Universe -- 390833306.
This project and SD were supported by the Baden-Württemberg Stiftung.



\bibliography{main}
\end{document}